# Phase noise-immune unconditionally secured classical key distribution using doubly coupled Mach-Zehnder interferometer

Byoung S. Ham

Center for Photon Information Processing, School of Electrical Engineering and Computer Science, Gwangju Institute of Science and Technology, 123 Chumdangwagi-ro, Buk-gu, Gwangju 61005, S. Korea

(Submitted on Sep. 06, 2020; bham@gist.ac.kr)

Recently, new physics for unconditional security in a classical key distribution (USCKD) in a frame of a double Mach-Zehnder interferometer has been proposed and demonstrated as a proof of principle, where the unconditional security is unaffected by the no-cloning theorem of quantum key distribution protocols. Due to environmental phase fluctuations caused by temperature variations, atmospheric turbulences, or mechanical vibrations, active phase locking seems to be necessary for the two-channel transmission layout of USCKD. Here, the two-channel layout of USCKD is demonstrated to be an environmental noise-immune protocol especially for free space optical links, where the transmission distance is potentially unlimited if random phase noises are considered.

**Introduction**

In quantum key distribution (QKD) technologies [1-15], unconditional security is provided by the no-cloning theorem of quantum mechanics using canonical variables, resulting from randomness via the Heisenberg uncertainty principle [16]. Randomness is the fundamental basis of unconditional security according to information theory [17]. In practice, however, quantum loopholes such as imperfect detectors and lossy quantum channels are major obstacles, resulting in conditional security as in classical technologies of algorithm-based protocols [9-15]. In addition to quantum loopholes, development of deterministic single photons or entangled photon-pair generators is far behind commercial implementations. Moreover, the deployment of long-distance quantum key distributions seems to not be possible due simply to the non existence of quantum repeaters [18]. In addition, the QKD key rate is extremely low compared with classical counterparts, and QKDs are not compatible with any classical systems in the real world such as fiber backbone networks and wireless mobile networks. Thus, implementations of QKD for practical applications may not be plausible in the near future. In other words, unconditionally secured information communications in the real world such as in nationwide on-line banking via quantum internet [19] are not possible with current QKD systems.

Since the recent investigation of quantumness regarding anticorrelation and photon bunching on a BS [20-27], a completely different physics for nonclassicality has been discussed [28-31] and experimentally demonstrated [32,33] for a coherence version of photonic de Broglie waves (CBW) and the unconditional security of classical key distributions (USCKD), where USCKD and CBW share the same relation as heads and tails of a coin [31]. This new idea of coherence quantumness is macroscopic based on bright coherent light with quantum superposition of two Mach-Zehnder interferometers via a specific coupling method. So far, quantum information has been limited to the microscopic world composed of a few atoms or photons, relying on the particle nature of duality [34]. The new interpretation about quantumness on a BS for photon bunching and anticorrelation, however, is based on the wave nature of photons, resulting in coherence quantum information, where such nonclassical features of anticorrelation, CBW, and USCKD cannot be obtained classically. For example, the physics of CBW lies in the controllable higher-order quantum superposition among independent Mach-Zehnder interferometers (MZIs) [27,32], where the coupling method plays a key role. Like a coupled two-mode pendulum model in classical physics [35], the quantumness in CBW has also been investigated using a tensor product of independent phase bases in coupled MZIs, where the phase bases satisfy the orthonormal conditions of a Hilbert space [31].

The heart of unconditional security in both QKD and USCKD is in the randomness of eavesdropping for the distributed keys. In QKD, the randomness is in Heisenberg's uncertainty principle of conjugate variables composed of two sets of orthogonal bases, where the no-cloning theorem provides the bottom line of no copying. On the other hand, in USCKD, the eavesdropping randomness of orthogonal bases is sought from the path superposition of MZI, where the realization of randomness is provided by key distribution determinacy between two remote parties via double unitary transformations [28,33]. Regarding USCKD, however, the two-channel layout of transmission lines should be vulnerable to environmentally-caused phase noises induced by temperature variations, mechanical vibrations, and/or atmospheric turbulences. It is well known that sophisticated phase control has been essential for MZI interferometers due to environmental phase noises [36-39]. Although state-of-the-art laser locking technologies are well implemented for various potential applications, active control for phase noises is still challenging and limits the maximum performance of the system. Here, we investigate inherent phase noise-immune characteristics of the double Mach-Zehnder interferometer in USCKD, where the phase noise-free characteristics can be applied for noise-limited information processing in both the wired- and wireless-based communications systems even without an active phase control.

**Results**

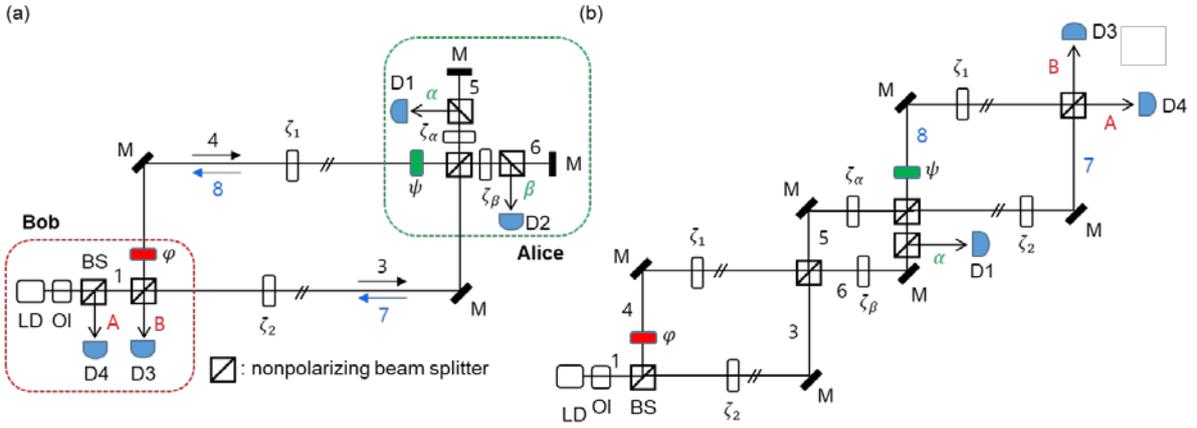

Fig. 1. A schematic of USCKD. (a) An original schematic of USCKD. (b) An unfolded schematic of USCKD. LD: laser diode, OI: isolator, BS: nonpolarizing beam splitter, D: detector, M: mirror. $\zeta_j$: environmental phase noise. The numbers indicate corresponding electric fields.

Figure 1 shows a schematic of the USCKD, where the environmentally-caused phase noises incurred in transmission channels of MZI are perfectly and automatically compensated via a round-trip transmission scheme. In Fig. 1(a), the shared transmission channels of MZI between two remoted parties, Alice and Bob, are not quantum but classical, where Bob controls the phase shifter $\varphi$ and detectors D3 and D4, while Alice controls the phase shifter $\psi$ and detectors D1 and D2. Here, 'classical' means that an eavesdropper can copy the light carrier in channels without revealing the eavesdropper's existence to both Alice and Bob, as allowed in classical cryptographic systems. The $\zeta_j$ represents environment-caused phase noise in each transmission channel, where $\zeta_1 \neq \zeta_2$ due to channel independency. The phase $\psi$ is for the light returned by Alice, where $\psi$ is invisible to the outbound lights '3' and '4.' This can be easily satisfied with an optical delay line inserted between two mirrors on Alice's side.

According to USCKD [30], Bob prepares a key with a random phase basis, $\varphi \in \{0, \pi\}$, where the phase bases are orthonormal to each other in the MZI system. Alice confirms Bob's prepared key with her phase basis choice, $\psi \in \{0, \pi\}$. Here, Alice's confirmation depends on a protocol, whether the confirmation can be limited to the same bases or all bases (see the Supplementary Information of ref. 30). If the bases are the same each other, the double unitary transformation results in an identity relation, otherwise an inversion relation for different basis combinations [30]. This two-channel key distribution process of USCKD is perfectly deterministic due to MZI directionality: If Bob's basis choice is for $\varphi = \pi$, Alice's detector D1 must click; If Bob's choice is $\varphi = 0$ and Alice's choice is $\psi = 0$, Bob's detector D4 clicks. The unconditional security is rooted in the superposition-caused measurement randomness in the shared MZI channels [30]. A sophisticated Eve, of course, can use the same measurement tool as Bob's or Alice's. However, the information extraction chance by Eve is 50% on average due to the indistinguishability in the superposed channels of MZI, resulting in measurement randomness. To avoid classical attacks such as memory-based attacks, an authentication or network initialization process is required [30].

Using matrix representations, the following analytic derivations are obtained for Fig. 1 (see Section A of the Supplementary Materials):

$$\begin{bmatrix} E_\alpha \\ E_\beta \end{bmatrix} = [MZI]_1 \begin{bmatrix} E_0 \\ 0 \end{bmatrix}$$

$$= \frac{1}{2} e^{i\zeta_1} \begin{bmatrix} e^{i\zeta} - e^{i\varphi} & i(e^{i\zeta} + e^{i\varphi}) \\ i(e^{i\zeta} + e^{i\varphi}) & -(e^{i\zeta} - e^{i\varphi}) \end{bmatrix} \begin{bmatrix} E_0 \\ 0 \end{bmatrix},$$

$$= \frac{1}{2} e^{i\zeta_2} \begin{bmatrix} 1 - e^{i(\varphi-\zeta)} & i(1 + e^{i(\varphi-\zeta)}) \\ i(1 + e^{i(\varphi-\zeta)}) & -(1 - e^{i(\varphi-\zeta)}) \end{bmatrix} \begin{bmatrix} E_0 \\ 0 \end{bmatrix}, \quad (1)$$

where $[MZI]_1 = \frac{1}{2} e^{i\zeta_2} \begin{bmatrix} 1 - e^{i(\varphi-\zeta)} & i(1 + e^{i(\varphi-\zeta)}) \\ i(1 + e^{i(\varphi-\zeta)}) & -(1 - e^{i(\varphi-\zeta)}) \end{bmatrix}$ and $\zeta = \zeta_2 - \zeta_1$. Thus, the effective channel noise $\zeta$ due to the environmental phase fluctuations results in:

$$I_\alpha = \frac{1}{2}[1 - \cos(\varphi - \zeta)]I_0, \quad (2)$$

Likewise,

$$I_\beta = \frac{1}{2}[1 + \cos(\varphi - \zeta)]I_0 I_0. \quad (3)$$

Thus, the output intensities of $I_\alpha$ and $I_\beta$ for a fixed $\varphi$ are time dependent, whose average is bounded by the classical limit of $I_0/2$. This means that the control of environmental phase fluctuations is critical in MZI interferometry [36-39].

For the returned light to Bob's side by Alice in Fig. 1, the matrix representations are as follows:

$$\begin{bmatrix} E_A \\ E_B \end{bmatrix} = [MZI]_2 [\zeta_{\alpha\beta}] \begin{bmatrix} E_\alpha \\ E_\beta \end{bmatrix}$$

$$= \frac{1}{4} e^{i(2\zeta_2 + \zeta_\alpha)} \cdot$$

$$\begin{bmatrix} -(1 - e^{i\zeta''})(1 + e^{i(\psi+\varphi-2\zeta)}) - (1 + e^{i\zeta''})(e^{i(\varphi-\zeta)} + e^{i(\psi-\zeta)}) & -i[(1 - e^{i\zeta''})(1 - e^{i(\psi+\varphi-2\zeta)}) + (1 + e^{i\zeta''})(e^{i(\psi-\zeta)} - e^{i(\varphi-\zeta)})] \\ -i[(1 - e^{i\zeta''})(1 - e^{i(\psi+\varphi-2\zeta)}) + (1 + e^{i\zeta''})(e^{i(\psi-\zeta)} - e^{i(\varphi-\zeta)})] & (1 - e^{i\zeta''})(1 + e^{i(\psi+\varphi-2\zeta)}) - (1 + e^{i\zeta''})(e^{i(\varphi-\zeta)} + e^{i(\psi-\zeta)}) \end{bmatrix} \begin{bmatrix} E_0 \\ 0 \end{bmatrix}$$

(4)

where $[MZI]_2 = \frac{1}{2} e^{i\zeta_2} \begin{bmatrix} 1 - e^{i(\psi-\zeta)} & i(1 + e^{i(\psi-\zeta)}) \\ i(1 + e^{i(\psi-\zeta)}) & -(1 - e^{i(\psi-\zeta)}) \end{bmatrix}$, $[\zeta_{\alpha\beta}] = \begin{bmatrix} e^{i\zeta_\alpha} & 0 \\ 0 & e^{i\zeta_\beta} \end{bmatrix}$, and $\zeta'' = \zeta_\alpha - \zeta_\beta$. Here $[\zeta_{\alpha\beta}]$ is the environmental phase noise occurred in the coupling MZI between two main MZIs of $[MZI]_1$ and $[MZI]_2$, and the phase noise of $\zeta''$ can be easily controlled technically.

(i)     For $\zeta'' = 0$

For $\zeta'' = 0$, equation (4) is represented as:

$$\begin{bmatrix} E_A \\ E_B \end{bmatrix} = \frac{-1}{2} e^{i(\zeta_1 + \zeta_2 + \zeta_\alpha - \varphi)} \begin{bmatrix} (e^{i(\psi-\varphi)} + 1) & i(e^{i(\psi-\varphi)} - 1) \\ i(e^{i(\psi-\varphi)} - 1) & (e^{i(\psi-\varphi)} + 1) \end{bmatrix} \begin{bmatrix} E_0 \\ 0 \end{bmatrix}. \quad (5)$$

Thus, the output intensities for $\zeta'' = 0$ are as follows:

$$I_A = \frac{1}{4}[1 + e^{i(\psi-\varphi)}][1 + e^{-i(\psi-\varphi)}]I_0$$

$$= \frac{1}{2}[1 + \cos(\psi - \varphi)]I_0. \quad (6)$$

$$I_B = \frac{1}{4}[1 - e^{i(\psi-\varphi)}][1 - e^{-i(\psi-\varphi)}]I_0$$

$$= \frac{1}{2}[1 - \cos(\psi - \varphi)]I_0. \quad (7)$$

Here, the relative phase noise $\zeta$ between outbound ('3' and '4') and inbound ('7' and '8') lights in the transmission channels of Fig. 1 can be different due to time delay, but assumed to be zero for simplicity or for a short distance or a slowly varying noise condition. Equation (5) satisfies the identity and inversion relation of the original protocol of USCKD with no influence of environmental phase noises if $\zeta'' = 0$, where the intermediate coupling MZI can be isolated from the phase noises (see the Supplementary Information).

(i)     For $\zeta'' \neq 0$

If the intermediate coupling MZI is exposed to phase noises of $\zeta_\alpha$ and $\zeta_\beta$, then the output fields of equation (4) are seriously affected by environmental noises in both $\zeta''$ and $\zeta$, resulting in lower visibility of identity or inversion. The related numerical calculations are shown in Fig. 2 as both functions of $\zeta''$ and $\zeta$ for $\varphi = \psi$ (see the case with $\varphi \neq \psi$ in the Supplementary Information).

For the numerical calculations in Fig. 2, the output field intensities are calculated for randomly varying $\zeta''$ and $\zeta$. Due to the random phase noises between 0 and $2\pi$, the output intensities of $I_A$ and $I_B$ result in random fluctuations between 0 and 1 in the unit of the input field intensity $I_0$. To analyze the results appeared in the upper panel, output intensities are averaged for all $\zeta''$ values at each random phase noise of $\zeta$ (see the middle panel). As discussed above, the output intensities are invariant to the random phase noise of $\zeta$ if $\zeta'' = 0$ (see the dotted lines of the left middle panel). Starting from these extreme values, the average values of the output intensity $I_A$ ($I_B$) gradually reduce (increase) from 1 (0) to 0.75 (0.25) as the random phase noise range of $\zeta''$ increases up to $\pi$. If the random phase variation range of $\zeta''$ increases more than $\pi$, the average of output intensities fluctuate up and down in a small range (see the colored curve bundle circled). These intensity fluctuations may be caused by the nondeterministic random values of $\zeta''$ affecting the output values in equation (4). The right middle panel shows ten repeated data for a fixed random phase range of $\zeta'' = 2\pi$, where each intensity variation should be reduced as the $\zeta''$ set increases (see the Supplementary Information). The interesting results is that the output field intensity values are clearly and completely separated each other and unexpectedly stable even under the random phase noises of both $\zeta''$ and $\zeta$. Because both intensity would vary between 0 and 1, the lowest (highest) average value of $I_A$ ($I_B$) should be $I_0/2$ in the average. As a result, the average value of the output intensities for all possible $\zeta''$ are expected to be a half of each maxima, resulting in $3I_0/4$ ($I_0/4$) for $I_A$ ($I_B$). Thus, USCKD is robust for the key distribution even under full scale of phase noises, where the detection bound for $I_A$ ($I_B$) can be set lower (higher) than for the optimum case. This is the environment-caused noise immune characteristics of USCKD.

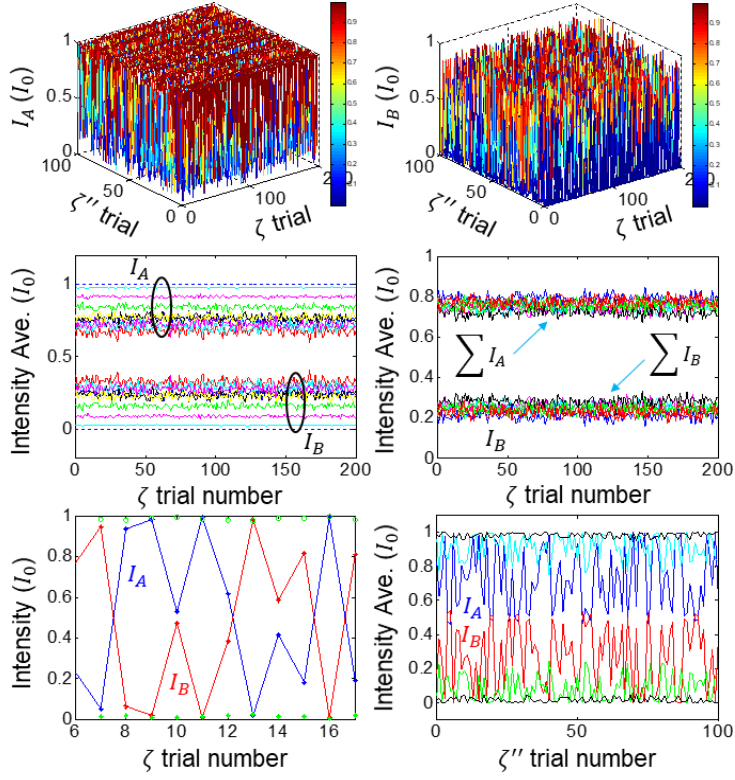

Fig. 2. Numerical calculations of equation (4) for $\zeta'' \neq 0$ and $\varphi = \psi = 0$. At each trial random phase is applied, where the r range of $\zeta''$ varies for a fixed range of $\zeta$. Top row: Individual output intensities for random phase variations of $\zeta''$ ($2\pi$) and $\zeta$ ($2\pi$). Middle row (left): Intensity average over $\zeta''$, whose random phase variation is $0 (dotted); \frac{\pi}{4}(cyan); \frac{\pi}{2}(magenta); \frac{3\pi}{4}(green); \pi(blue); \frac{5\pi}{4}(red); \frac{3\pi}{2}(black/yellow); \frac{7\pi}{4}(green); 2\pi(red)$. Middle row (right): Repeated (10 times) trials for the range of $\zeta'' = 2\pi$. Bottom row (left): Individual intensities for the Middle right: blue/red lines indicate $I_A/I_B$ for $\zeta'' = 19$; green circle/star for $\zeta'' = 23$. Bottom row (right): Intensity average over $\zeta$, where the random phase variation of $\zeta''$ is $2\pi$ (red & blue), $\pi/2$ (green & cyan), and $\pi/5$ (black) for maximum variation of $\zeta$.

The left bottom panel of Fig. 2 shows both output intensities for individual $\zeta''$ by expanding one in the right middle panel. The individual full fluctuations in both output fields seems to be obvious due to the random phase noises. Thus, averaging effect is extremely important for the intensity stabilization of the output fields. Unlike the program setting with random phase noises for a long time period, however, environmental noises should fluctuate continuously in a short time scale far less than a MHz rate. In Fig. 1, the phase noise variation of $\zeta''$ should be much slower than that of $\zeta$ due to much shorter path lengths. Thus, it is more important to understand output fields relation with respect to $\zeta''$.

The right bottom panel of Fig. 2 shows both averaged output intensities over the transmission channel noise $\zeta$ as a function of $\zeta''$ for different phase noise ranges. The blue and red curves are for the upper panel with a range of $\zeta'' = 2\pi$. Like the middle panel, the intensity fluctuations of the output fields are still less than 1/2 with hardly crossing over the middle line, resulting in a safe separation between $I_A$ and $I_B$. As the random variation range of $\zeta''$ decreases from $\zeta'' = 2\pi$ to $\zeta'' = \pi/5$, the $\zeta-$averaged intensity fluctuations are greatly reduced to nearly zero (see the black curves). Thus, the USCKD scheme of Fig. 1 is safe to the environment-caused random phase noises, where $\zeta''$ is a technical matter.

Figure 3(a) shows the related experimental results for the bottom panel of Fig. 2, where the $\zeta''$ effect on the output fields is critical in a slow variation of the phase noise as analyzed in Fig. 2. The fairly $\zeta''$−noise-subdued USCKD has already been observed in ref. 33 for the same setup of Fig. 1. Here, there is no active phase noise control in Fig. 3(a), but the setup is just put in a quiet, calm, and closed room. In Fig. 3(a), the output fields are averaged by 10 times internally by a Tektronix digital oscilloscope. Under the common lab condition, the output fields slowly fluctuate and cross slightly over the half line (see the dotted circle) each other. This slight cross-over has been shown in the bottom panel of Fig. 2.

In Fig. 3(b), such a slight crossover observed in Fig. 3(a) is analyzed with different average number n over the random $\zeta$−caused intensity values for a specific value of $\zeta''$, where $\zeta''$ is replaced by a linear function $\zeta'' \in \{0, 2\pi\}$. For this, two different average numbers (n=20; 2,000) are compared each other for the $\zeta$−averaged output intensities. As analyzed in equations (4)-(7), the output intensities are immune to the phase noise if $\zeta'' = 0$, where this analysis is numerically confirmed at both ends of Fig. 3(b). For $\zeta'' \neq 0$, Fig. 2 shows random phase noise-dependent intensity fluctuations. Unlike Fig. 2, in Fig. 3(b), we deal with averaging-dependent intensity fluctuation as a linear function of $\zeta''$ with no noise variation. For n=2000, the output intensities are fairly stable regardless of $\zeta''$ (see the green and cyan curves). For n=20, however, the output field fluctuations gradually increase as $\zeta''$ increases from 0 to $\pi$ and then reduced down symmetrically. Here, $\zeta'' = \pi$ stands for a maximum noise to the system according to equation (4). Thus the crossover in Fig. 3(a) is explained with the low averaging number in Fig. 3(b). This result is quite interesting because an active phase control has been intensively sought in interferometry [36-44]. Both high number of averaging in Fig. 3(b) and low noise range of $\zeta''$ in the right bottom panel of Fig. 2 give a great potential not only for USCKD benefit but also for general benefit to conventional interferometry especially for the highly stable locking system including gravitational wave detection.

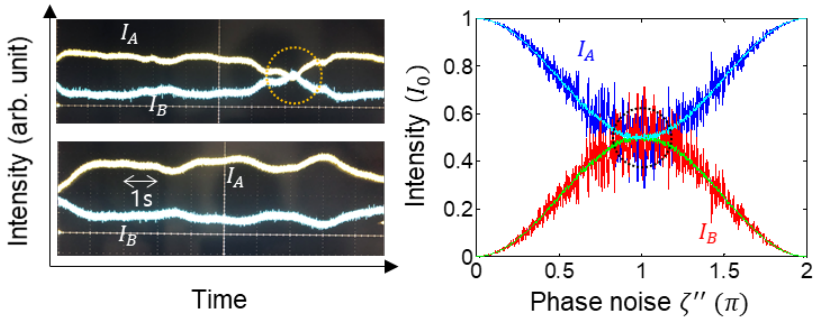

Fig. 3. Experimental demonstrations of equation (4) for $\zeta'' \neq 0$. (Left) Experimental results for the average of n=10. (Right) Numerical analysis for random phase noise variation of $\zeta''$ with averaging for n=20 (blue and red) and n=2,000 (cyan and green).

**Discussion**

Unlike the one-way MZI system of equations (1)-(3) [38,39], equation (4) is for the round-trip MZI system of Fig. 1 representing the environmental noise effect on USCKD. As analyzed in equations (5)-(7), the transmission channels of the shared MZIs of USCKD satisfies environmental noise-immune property by inherent self-noise cancellations if $\zeta'' = 0$. Because the intermediate coupling MZI for $\zeta'' = 0$ is easily obtained by the state-of-the-art laser locking technologies, the output intensity noise fluctuations of the USCKD can be fairly subdued no matter how big the channel noise is as shown in Fig. 2. Even with full random noise in the intermediate MZI, both averaged output intensities are pretty stable and fairly separated to be applied for the key distribution of USCKD. If the environmental noise varies slowly so that large averaging is not possible for fast key distribution process comparable to the round-trip travel time, then the output values slightly cross over the half line. Instead of $\zeta''$ noise isolation by an active or passive control, intentionally added random phase noise would be a rough solution. Moreover, well random noise averaged output intensities show a distinct separation between two regions. As a result, the schematic of USCKD in Fig. 1 suffices the environmental noise-immune system for both fast and slow noise environments. This environmental phase noise-immune property of the present USCKD system has a novel feature especially for the free-space key distribution, where atmospheric fluctuations have been a critical obstacle for the optical link in both quantum [41-43] and classical [44,45] communications.

**Conclusion**

We analyzed, discussed, and experimentally demonstrated the environmental noise-immune USCKD protocol in a coherence scheme of classical physics, where the phase noise to the long pair of transmission lines of MZI is inevitable in practice. The noise immune USCKD is due to self-compensated phase noise in the round-trip MZI configuration, where the round-trip scheme is for the identity and inversion relation via double unitary transformation. However, the phase fluctuations in the intermediate coupling module between the inbound and outbound MZI channels were serious to demolish the USCKD protocol if the noise is slowly varying or the processing is fast enough not to average fair numbers. Such a phase noise can also be passively controlled by inserting on-purpose random noises to the system. Although perfect noise immune USCKD can be reached with zero phase noise in the intermediate coupling MZI, such purity of USCKD can also be a defect to the diagnosis of the transmission lines from attacks. Based on free space optical links limited by a few km distance, the demonstrated phase noise-immune USCKD configuration shows a great potential for ground-to-satellite optical links as well as ground-to-air optical links. Thus, the present demonstration may open the door to a new realm of secured information communications beyond both classical and quantum security.

**Methods**

Numerical calculations: In Figs. 2 and 3, home-made Matlab programs are used for numerical claculations, where the random phase noise is obtained from the rand(1) commend. In Fig. 2, various phase noise range was controlled by multiplying a certain number to the output of rand(1).

Experiments: In Fig. 3(a), the wavelength of laser is 606 nm from Toptica TA-SHG pro, but not limited. The optical power of the input light is ~1 mW. After division by the first beam splitter in Fig. 1, each channel was controlled by each AOM driven by synchronized rf generators (PTS 160/250; Tektronix AFG3102), resulting in same initial phase (see ref. 29 for details). The data was captured in the screen of an oscilloscope (Tektronix DPO 5204B) via avalanche photodiodes (Hamamatsu C12703).

**Acknowledgments**

The present work was supported by the GRI grant of GIST in 2020.